\documentclass[draft]{agujournal2019}
\usepackage{url} 
\usepackage{lineno}
\usepackage{comment}
\usepackage[inline]{trackchanges} 
\usepackage{soul}
\usepackage{amsmath}
\usepackage{array}
\usepackage{graphicx}
\usepackage{bm}
\usepackage{accents} 

\usepackage{subcaption}
\draftfalse

\begin{document}

\title{Probabilistic Solar Proxy Forecasting with Neural Network Ensembles}


\authors{Joshua D. Daniell$^{1}$ and Piyush M. Mehta$^{1}$}
\affiliation{1}{Dept. of Mechanical and Aerospace Engineering\\
West Virginia University
Morgantown, WV 26505}

\correspondingauthor{Joshua Daniell}{jddaniell@mix.wvu.edu}

\begin{keypoints}

\item Ensemble methods show promise for short term forecasting of $F_{10.7}$ and are able to provide robust and reliable uncertainty estimates.

\item Neural network (non-linear) methods provide better short term forecasting than linear auto regressive methods.

\item Addition of input data by backwards averaging or varying lookback window improves model performance.

\end{keypoints}

\begin{abstract}
Space weather indices are used commonly to drive forecasts of thermosphere density, which directly affects objects in low-Earth orbit (LEO) through atmospheric drag. One of the most commonly used space weather proxies, $F_{10.7 cm}$, correlates well with solar extreme ultra-violet (EUV) energy deposition into the thermosphere. Currently, the USAF contracts Space Environment Technologies (SET), which uses a linear algorithm to forecast $F_{10.7 cm}$. In this work, we introduce methods using neural network ensembles with multi-layer perceptrons (MLPs) and long-short term memory (LSTMs) to improve on the SET predictions. We make predictions only from historical $F_{10.7 cm}$ values, but also investigate data manipulation to improve forecasting. We investigate data manipulation methods (backwards averaging and lookback) as well as multi step and dynamic forecasting. This work shows an improvement over the baseline when using ensemble  methods. The best models found in this work are ensemble approaches using multi step or a combination of multi step and dynamic predictions. Nearly all approaches offer an improvement, with the best models improving between 45 and 55\% on relative MSE. Other relative error metrics were shown to improve greatly when ensembles methods were used. We were also able to leverage the ensemble approach to provide a distribution of predicted values; allowing an investigation into forecast uncertainty. Our work found models that produced less biased predictions at elevated and high solar activity levels. Uncertainty was also investigated through the use of a calibration error score metric (CES), our best ensemble reached similar CES as other work.
\end{abstract}
\section*{Plain Language Summary}
$F_{10.7}$ is a radio wave measurement that closely follows solar activity. Increased solar activity can cause heating and density changes in the Earth's upper atmosphere. These density changes directly affect the drag force felt by satellites and can alter their orbit. In order to better predict the upper atmosphere density and paths taken by these spacecraft, we need to increase our prediction skills of $F_{10.7}$. In this work, we use machine learning techniques to make a set forecasts of $F_{10.7}$ that can be combined to provide less error than currently used forecasting methods. Forecasting through these ensemble methods outperform a commonly used operational method. In addition, this set of predictions can be used to determine a level of confidence in the forecasts.

\section{Introduction}
The number of objects in the low-Earth orbit (LEO) region is growing quickly, especially with the work done by the private sector, such as SpaceX and OneWeb. The injection of many objects into LEO, especially by satellite constellations, necessitates better space situational awareness for future object location as well as trajectory. Objects in LEO are affected by many orbital perturbations, but atmospheric drag accounts for the largest source of uncertainty. Atmospheric drag is directly tied to thermosphere heating and has been historically correlated to the $F_{10.7cm}$ radio flux proxy. As the number of objects grows, it is necessary for the scientific community to more robustly model and predict the space environment as well as the dynamics of Earth's atmosphere. 


The properties of Earth’s upper atmosphere are heavily impacted by solar activity, specifically extreme ultra-violet (EUV) irradiance. Atmospheric density changes occur most often with solar activity level. The high energy EUV solar radiation is absorbed by the Earth’s upper atmosphere and causes large density variations due to heating. The change in atmospheric density due to solar EUV directly impacts the dynamics of LEO objects in the thermosphere. More robust predictions for solar activity will lead to more robust predictions of satellite dynamics in LEO. Due to the atmosphere’s opaqueness to solar EUV, we cannot measure EUV directly on the ground, but we can measure $F_{10.7 cm}$. Historically, the $F_{10.7 cm}$ solar radio flux proxy has shown high correlations with solar EUV. 

The $F_{10.7 cm}$ solar radio flux proxy, denoted as $F_{10.7}$ for the remainder of the work, is one of the most widely used proxies for solar activity. \citeA{Tapping2013} describes the proxy measurement as a “determination of the strength of solar radio emissions in a 100 MHz-wide band centered on 2800 MHz (a wavelength of 10.7 cm) averaged over an hour”. The $F_{10.7}$ proxy is reported in solar flux units (SFU), where
\begin{center}
    1 $SFU = 10^{-22}\frac{W}{Hz  m^2}$
\end{center}
The $F_{10.7}$ proxy has a high correlation with both sunspot number and solar EUV irradiance, seen in both \citeA{Svalgaard2010} and \citeA{Vourlidas2018}; and is considered an indirect measure of solar activity levels. Care should be taken to describe the difference between index and proxy; $F_{10.7}$ has been found to be highly correlated to solar activity \citeA{Wit2017} yet is not a direct measurement of solar EUV. Similarly to \citeA{Licata2020} we refer to $F_{10.7}$ as a proxy for solar activity and not an index, since we are using a measure of radio flux that is correlated with solar EUV.
\subsection{Currently Used Models}
Due to the relation between $F_{10.7}$ as a solar EUV forcing term of the upper atmosphere and neutral thermosphere density modeling, more robust predictions of the $F_{10.7}$ proxy would, in turn, provide better modeling and prediction of density in the thermosphere. Commonly used prediction methods for the $F_{10.7}$ proxy using historical values include a mixture of statistical, autoregressive, linear, and deep learning methods. These varied methods have produced a variety of conclusions, some models claim that linear methods outperform neural networks seen in \cite{Warren2017} and \cite{Huang2009}. Other authors have claimed that they can use machine learning methods to outperform linear methods like \cite{Stevenson2022} and \cite{Luo2022}. Methods, such as these, must be carefully compared in order to reduce contradictory conclusions.
\subsubsection{Linear Methods}
The persistence model is usually considered a naïve model and is a first step used to compare the performance of various time series forecasting methods. To model persistence, the previous time step value is persisted up to the desired forecasting horizon, \textit{H}, where $t$ is the day a forecast is made. 
\begin{equation} 
   F_{10.7_{t}}= F_{10.7_{t+1}}=F_{10.7_{t+2}}=...=F_{10.7_{t+H}}
\end{equation}

In order to forecast the $F_{10}$ proxy, SET uses a linear prediction algorithm that captures recurrence and persistence. The algorithm that is used is the “TS\_FCAST” subroutine in the Interactive Data Language (IDL).  This linear prediction algorithm fits and uses a \textit{pth} order auto regressive model on \textit{p} days lookback. The model is re-fit on data every time a prediction is to be made.
\begin{equation}
    x_t = \phi_{1}x_{t-1}+\phi_{2}x_{t-2}+...+\phi_{p}x_{t-p}+w_{t}
\end{equation}

The short term (6-day) predictions made by this method have been benchmarked and analyzed thoroughly by \citeA{Licata2020} and will be considered as the baseline model for the comparisons of the $F_{10}$ predictions made in this work. This algorithm uses auto regression and relies on a predictive method called “dynamic forecasting” for which one-step predictions are calculated recursively to reach the prediction multiple time steps ahead of the forecast epoch.

According to \cite{Warren2017}, a similar method to the baseline was used for prediction of $F_{10}$ using a multi step linear prediction. The authors used a linear regression method for predicting values at multiple time steps using a set of various linear regression coefficients calculated for each horizon day. The authors also compared the use of artificial neural networks for prediction of $F_{10.7}$ with that of the linear model. The authors concluded that “forecasting via sophisticated artificial neural networks is not any better than a simple linear forecasting approach”. An additional method by \cite{Wang2018}, performs multi step prediction using linear methods which considers both the correlation between predicted time steps as well as the heteroscedastic properties. 
\subsubsection{Non-Linear Methods}

\cite{Huang2009} investigated the usage of a Support Vector Regression (SVR) model to perform short term (3 day) predictions of $F_{10}$. An SVR approach uses a non-linear mapping of input data into a higher dimensional feature space, which is then fitted via linear regression. This linear regression on the higher dimensional space is optimized via minimization of a cost function.  The authors determined that that “our approach can perform well by using fewer training data points than the traditional neural network.” 

\cite{Stevenson2022} discussed an approach for $F_{10}$ forecasting using neural networks. The authors implemented the network architecture created by \cite{Oreshkin2019}, Neural Basis Expansion Analysis for Interpretable Time Series Forecast (N-BEATS), for multi step forecasting proxy values up to a horizon of 27 days. N-BEATS uses fully connected layers linked to non-linear basis functions to make both a forecast and back cast while training. 

N-BEATS was proposed for usage without the need for knowledge of the domain. \cite{Stevenson2022} used a neural network ensemble approach suggested by \cite{Oreshkin2019} for the direct predictions of $F_{10}$ and associated uncertainty. N-BEATS was also used due to the computational efficiency over typical recurrent neural networks. It was determined that the NBEATS ensemble approach “systematically outperformed” both the statistical British Geological Survey (BGS) approach and the persistence model. The N-BEATS ensemble also achieved improved or similar performance when compared to the CNES (French Space Agency) CLS model, which is a shallow neural network based on 4 different radio flux wavelengths.

\cite{Luo2022} acknowledge the previous methods for forecasting $F_{10}$ rely heavily on linear methods and propose the usage of Convolutional Neural Networks (CNNs) and LSTMs. The usage of linear methods for forecasting $F_{10}$ results in relatively stable mid to long term predictions but fall short when high quality predictions on a smaller horizon are required. Luo et al. also proposes that a CNN will be of use in extracting features of the $F_{10}$ time series, and an LSTM will be useful in prediction of future values. Luo et al. considered forecasting values of the adjusted proxy, instead of the observed proxy and the authors found an improvement over the linear methods by using ML methods.

Some models for forecasting $F_{10}$ rely on inputs other than previous observations such as \cite{Huang2009}, who proposed the addition of a flux tube expansion factor, $f_{s}$, to the input of the SVR prediction method. The authors proposed that by including this as an input, the prediction of $F_{10}$ may be improved during large spikes in solar activity. The authors acknowledge that the SVR method is general and could use improvement, and that the addition of an input that represents general solar magnetic activity may help $F_{10}$ prediction, due to subtle changes in the solar magnetic field.

\cite{Benson2021} used an LSTM model to forecast various solar proxy and geomagnetic indices simultaneously. The authors considered the interaction between $F_{10}$, $F_{30}$, $F_{15}$, geomagnetic indices, and solar imaging in forecasting of proxy/index values one solar rotation (27 days) in advance. The authors concluded that machine learning methods outperform linear regression, persistence, and mean value models. It was also determined that the addition of solar imaging data improves predictions when compared to proxy/index values only.

Forecasting of $F_{10}$ using traditional methods, such as the baseline, provides a deterministic forecast, which prevents full analysis on uncertainty of prediction. A key issue that needs to be addressed in the field of forecasting $F_{10.7}$ is uncertainty in the predicted value. By providing a probabilistic forecast, the resulting proxy can be used for uncertainty analysis in thermosphere density modeling and steps may be taken in operations with the knowledge of forecast uncertainty, such as best or worst case situations.
This work builds off previous machine learning methods and implements neural network ensembles, or a set of individual predictions, to improve on linear methods while also providing probabilistic forecasts which can be sampled for operation.

The work is organized as follows. In Section 2, we introduce the $F_{10}$ data set and data preparation procedures. In section 3, we discuss popular neural network types, inputs and outputs, and neural network ensembles. We discuss how these models can be implemented for $F_{10}$ forecasting. In Section 4, we compare the results of previously used models for short term forecasting of $F_{10.7}$ with the introduced neural network ensemble and single model methods we discuss the benefits and drawbacks of such ensemble methods and the distributions they produce. We also discuss the uncertainty in predictions as well as performance at various solar activity levels.


\section{Data}
The $F_{10.7}$ proxy has been recorded consistently since 1947 by sites in Canada. Originally, these observations were performed by a site south of Ottawa, Ontario. Due to expansion of city infrastructure, large amounts of disturbances were noted in observations and, in 1962, a new observation site was constructed at the Algonquin Radio Observatory. A secondary flux measurement station was established at the Dominion Radio Astrophysical Observatory (DRAO). Since the establishment of DRAO, both the Algonquin and Ottawa sites have closed, and measurements are made solely at DRAO \cite{Tapping2013}.
Daily observed values of the Penticton $F_{10.7}$ proxy have been archived. This data set, recorded since 1947, has many missing values during the first years of operation and will be omitted. The data set that will be used begins on 1/1/1948 and runs through 12/10/2021, and can be seen in Figure \ref{fig:TrainTestSplit}. Dates for which daily flux values are missing, have been replaced with the last observed value. Forecasting of $F_{10.7}$ varies in skill with solar activity level as cycles over an 11 year period and various activity levels are encountered during forecasting periods.

\subsection{Preparation of Data}
\subsubsection{Data Normalization For Neural Network Training}
There have been several investigations that have performed space weather proxy forecasting using both single input and multiple input variables. These inputs usually contain auto-regressive (AR) information, represented as previous observations of the proxy. In \cite{Stevenson2022} the authors were able to improve upon the persistence baseline and other operational models, using only previous $F_{10}$ values. In this work, we limit ourselves to forecasting of the proxy using only historical data, to directly compare the ability of forecasting using neural network ensembles with the baseline method.

Neural networks require careful data preprocessing to effectively train the model and limit exploding or vanishing gradient problems, which occur in differentiation steps. \cite{Sola1997} showed that data normalization is crucial to obtain both good results and decrease computational time. This can be done with the standard normalization equation, given as
\begin{equation}
    \widetilde{X} = \frac{X-mean(X)}{standard \; deviation(X)}
\end{equation}
\subsubsection{Sliding Window Method}
To reduce the data set down to a form that can be learned by the neural network, one must split the data into input and output (or target) pairs. For time series forecasting, it is common to consider both lookback and forecast horizons to split the data.
Lookback, L, represents the past observations to include into a sample for the model. 
\begin{equation}
    Lookback\;Values = [F_{{10}_{T-1}},F_{{10}_{T-2}},F_{{10}_{T-3}}, ..., F_{{10}_{T-L}}] 
\end{equation}

The lookback can have a significant effect on the predicted value by a model. A larger lookback may allow a long-term trend to be identified by the model but may hinder the models’ performance due to lack of short-term trends that would be seen. It is necessary to investigate the effects of various lookbacks on the performance of a model and it may be beneficial to combine various lookbacks.
To be consistent in the analysis with the original benchmarking of forecasts done by \cite{Licata2020} on the SET algorithm, we consider a short term prediction $H$ = 6 Days, which will facilitate direct comparison. 
\begin{equation}
    Horizon \;Values = [F_{{10}_{T}},F_{{10}_{T+1}},F_{{10}_{T+3}}, ..., F_{{10}_{T+H}}] 
\end{equation}

\begin{figure}[!h]
    \centering
    \includegraphics[scale=0.4]{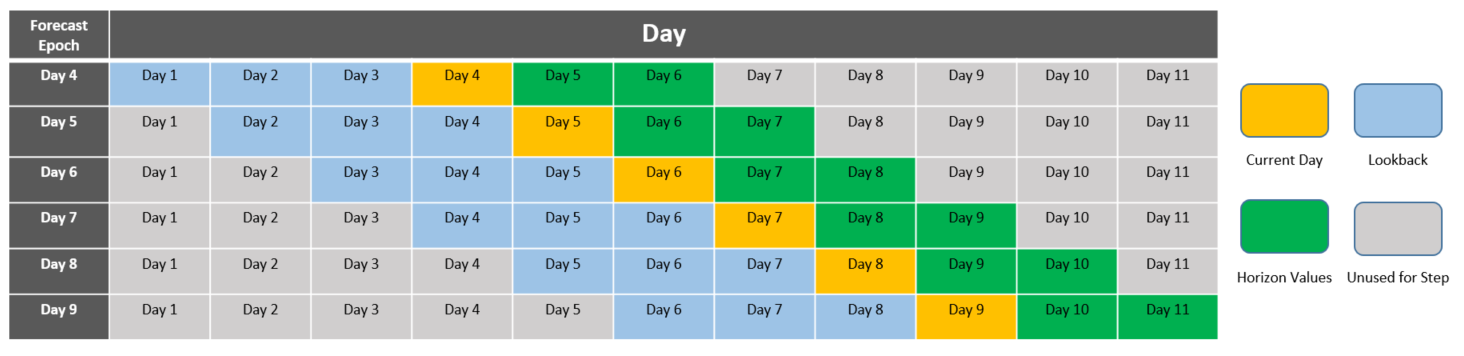}
    \caption{An example sliding window approach using a lookback of $L=3$ days and a horizon of $H =2$ days, with predictions starting on Day 4}
    \label{fig:Sliding Window Approach}
\end{figure}

The sliding window approach, seen in Figure \ref{fig:Sliding Window Approach} is used to separate the data into input/output pairs, called samples. The sliding windows creates a set of samples. The samples created have input of size (Input Features x $L$) and output of size 1x$H$. In the case of this work, we only consider the forecasting of a single feature, $F_{10}$.

We limit ourselves to using only $F_{10.7}$ values and trend knowledge in our model, but we explore data augmentation to provide the model with more information to learn from. Specifically, we investigate including an average historical value over a short-term period (backwards averaging). This can be done to provide a model with somewhat of a short term trend directly and may help with prediction of more stable values.

\cite{Licata2022a} discussed the necessary data preparation for LSTM model training. A standard feed forward NN requires the samples to have the same length about the first axis to achieve supervised training. “Consider the number of inputs ($n_{inp}$) and number of outputs ($n_{out}$)) for a given data set with n samples. The concatenation will result in an array of shape n × ($n_{out}$ + $n_{inp}$). The data must be stacked, so each row contains outputs and inputs for each lag-step ($n_{LS}$) and the current step. This orders in least-to-most recent from left-to-right. The data will now be of the shape n × ($n_{LS}$ + 1)($n_{out}$ + $n_{inp}$). The last $n_{inp}$ columns are then dropped as they are not needed. The data can be split into training inputs and outputs where the first n × $n_{LS}$($n_{out}$ + $n_{inp}$) columns are inputs and the last $n_{out}$ columns are the associated output. The final step is to reshape the input data to the shape n × $n_{LS}$ × ($n_{out}$ + $n_{inp}$).”

\subsubsection{Separation of Data into Train, Validation, and Test Sets}
To evaluate the performance of a neural network model, care must be taken to split the data into various sets known as test, training, and validation. The training set is the portion of the data that is fed into the network, for which weights and biases are adjusted to fit the data. Network weights are updated over training epochs in order to minimize a loss function (i.e. fitting the data.) Typically, in regression, the MSE loss is considered.
The validation set is a subset of the training set, that provides data to evaluate the performance of a model on an unseen set. By incorporating a validation set, model types can be compared on an unseen set without any leakage of testing data. The test set should be a set that is completely unseen during training. This is done to evaluate the true performance of a model on a set with no prior knowledge.
\begin{figure}[h]
    \centering
    \includegraphics[scale=0.5]{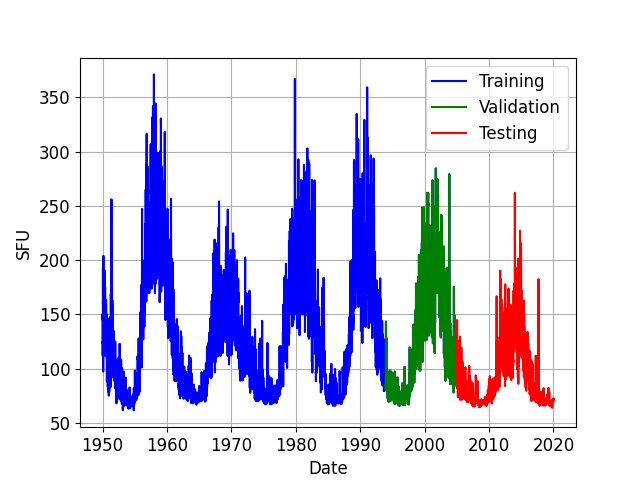}
    \caption{Training, validation, and test data are sampled to be consistent with prior work.}
    \label{fig:TrainTestSplit}
\end{figure}
\begin{table}
\label{Splittage}
\caption{Splitting of the Dataset}
 \centering
 \begin{tabular}{l c c }
 \hline
  Set  & Dates (YYYY/MM/DD) & Percentage of Total Data \%\\
 \hline
   Training  & 1950/01/01 - 1993/12/12 & 63\\
   Validation & 1993/12/13 - 2004/12/31 & 16 \\
   $^{a}$Testing  & 2005/01/01 - 2020/01/01 & 21  \\
 \hline
 \multicolumn{2}{l}{$^{a}$Contains sets used by both [7] and [4] for consistency.}
 \end{tabular}
 \end{table}
It is desired to split the data in such a way as to include a full solar cycle in the validation data; this is done to allow all four solar activity levels to be validated during the training process. By including partial solar cycles, the model would be missing validation data on certain solar activity levels and would be expected to under perform when those activity levels are seen during testing or operation.

For comparison against other methods, we carefully consider the process of splitting training, testing and validation data. Testing sets to be used contain both the test set used by N-BEATS \cite{Stevenson2022} and the benchmarking of the baseline model discussed by \cite{Licata2020}. Using the splitting seen in Figure \ref{fig:TrainTestSplit}, the test set will allow for the $F_{10}$ ensemble prediction performance presented in this work to be compared on the same test data used by both N-BEATS and the baseline. 

\section{Methodology}
\subsection{Neural Networks (NNs)}
In the past few decades, there has been a large amount of research in the usage of machine learning models known as Artificial Neural Networks (ANNs) to solve classification and regression tasks. Using a similar idea to neurons and linkages in the human brain, an ANN is used for a universal function approximation which is considered a supervised learning approach using nonlinear functions. A typical usage for regression ANN models is to solve for future time series values, known as prediction.

One of the most common neural network model types is the Multi-Layer Perceptron (MLP). Introduced by \cite{Rosenblatt1958}; an input parameter or set of input parameters is introduced into a layer of neurons. An activation function is applied to inputs (and weights in further layers). The final layer values pass through, in the case of regression, a linear activation function and are output from the model.

Another commonly used model for time series forecasting is Long-Short Term Memory (LSTM), which was introduced by \cite{Hochreiter1997}. LSTM is a modification of the traditional Recurrent Neural Network (RNN), which prevents the vanishing/exploding gradient problems that were seen in RNNs. LSTM leverages previously seen time series values stored in the LSTM cell’s hidden state “short-term memory”, to make a prediction more skillfully.

During training, the LSTM learns not only the short term trends and how to predict data, but also the relevance of previous information even with large time lags \cite{Hochreiter1997}. This makes them versatile and capable of picking up general trends. In this application, the LSTM can use its skills to better predict the underlying physics, allowing it to outperform the baseline model. The LSTM uses a \textit{hidden state} to "remember" previous predictions. 

\subsection{Multi-Step vs One Step (Dynamic) Forecasting}
\begin{figure}[h]
    \centering
    \includegraphics[scale=0.4]{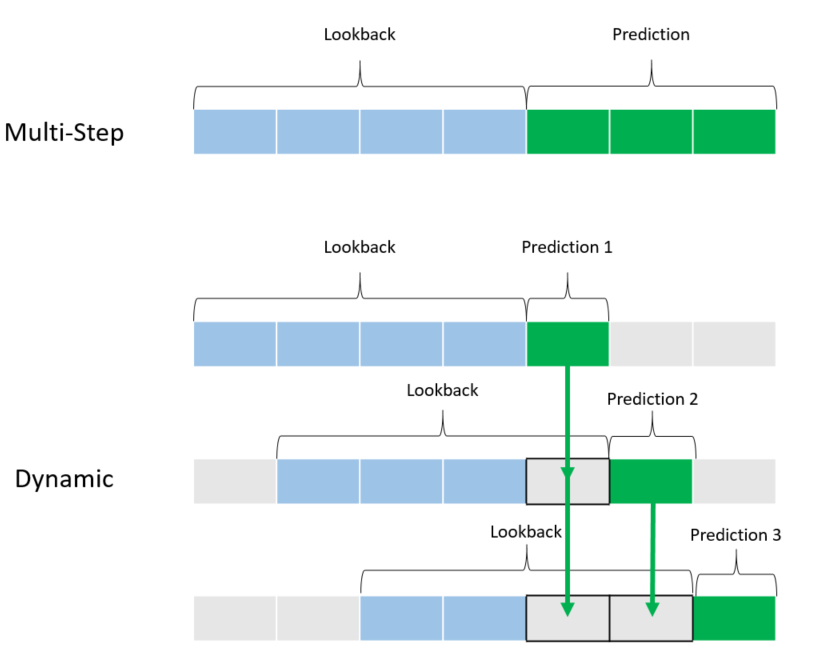}
    \caption{Consider a lookback of $L$ = 4 days, we can predict a horizon of $H$ = 3 days either by predicting all days at once (Multi-Step) or by recursively predicting single steps 3 times.}
    \label{fig:Dynamic Method}
\end{figure}

When splitting the data set using the sliding window approach, the outputs are of size $H$, and a model is directly trained to predict values for a horizon of $H$ days. \cite{Marcellino2006} concluded that dynamic forecasting typically outperform direct forecasts.
To explore dynamic predictive methods, consider setting the training targets to a size of 1, teaching the model to predict the next day. The model training is limited to prediction of directly the next day and may perform very well at predicting one day rather than attempting to generalize multiple days at once. To create a “Dynamic Forecast”, we perform a single step prediction, shift the input by a time step, and append the predicted value to the new input. This recursive method is done $H$ times to generate our forecast, illustrated in Figure \ref{fig:Dynamic Method}. A benefit of this dynamic forecasting approach is the ability to actively change the prediction length. The model is trained for one day at a time but operationally could be used for any desired horizon, without the need for further training.
\subsection{Neural Network Model Ensembles}
Typical probabilistic forecasting via machine learning methods involves generating a distribution of values for each time step, essentially generating both a mean and variance of future time steps. With this method, inherent assumptions of the distribution of future values are made during training or sampling, such as the assumption of a specific distribution, like Gaussian or Poisson. Extensive work in the field of machine learning has led to advancements in deterministic forecasting but due to limitations encountered in the learning process, a single machine learning architecture may not perform well in all areas of forecasting. For example, a model may be more skilled at predicting $F_{10}$ at a low solar activity level or may provide a constant biased prediction. By leveraging the concept of neural network ensembles, a combination of various models may improve prediction when compared to even the best performing single model. In addition to an improvement of direct value forecasting, an ensemble approach will create a distribution of predicted values, for which statistics can be evaluated and an uncertainty analysis can be performed. 

Due to the stochastic nature of the training process when weights are initialized randomly, different local minima may be encountered. \cite{MendesMoreira2012} states that with regards to ensemble generation, “a successful ensemble is one with accurate predictors that makes errors in different parts of the input space.” The authors further discuss the problem of combining ensemble predictions, or ensemble integration, and a process for removing models for which redundancy occurs. The ensemble process can be reduced to three main steps \cite{GIACINTO2001699} and is referred to as the overproduce and choose method. The steps are ensemble generation, ensemble pruning, and ensemble integration. Generation is the first step of this method and involves creating a set of models that attempt to fit the data. Pruning involves eliminating some of the models generated in the first step, usually models that provide redundant predictions. Integration involves determining the best way to combine these models. Using these 3 steps, a set of models can be transformed into an ensemble.
\subsubsection{Ensemble Diversity}
A difficult question is posed when generating an ensemble, “which models should be used?” This can be addressed by a key concept in model ensembles known as \textit{diversity}. Model diversity \cite{MendesMoreira2012} is defined as the study of the degree of disagreement between models. Another method of ensemble generation, discussed by \cite{Liu2000}, involves using interaction between ensemble members during training, in an evolutionary ensemble.

For this case, we choose to generate our model members independently, to compare results to \cite{Stevenson2022}, then proceed with integration. Several methods for promoting diversity in model generation are discussed by \cite{Sagi2018}.
\begin{itemize}
    \item \textbf{Input Manipulation:} Base models are fitted using different input data.
    \item \textbf{Manipulated Learning Algorithm:} Manipulation of the way each base model traverses the hypothesis space. In the case of neural networks, this can be achieved by varying hyper parameters, weights, optimizer, etc.
    \item \textbf{Partitioning:} Dividing the original training set into subsets then using each subset to train different models.
    \item \textbf{Ensemble Hybridization:} A combination of 2 or more of the other methods when generating an ensemble.

\end{itemize}
The authors also discuss methods for combining outputs of the ensemble members in the integration step.
Ensemble members can be combined using various weighting methods. In the most basic form, one could take a simple average of ensemble members,
\begin{equation}
    \overline{\hat{y}} = \frac{1}{M}\sum_{m=1}^{M} \hat{y}_m
\end{equation}
where $M$ is the number of ensemble members used, $\hat{y}_m$ is the prediction made by the $m^{th}$ model, and $\overline{\hat{y}}$ is the ensemble prediction.

\begin{figure}[h]
    \centering
    \includegraphics[width=\textwidth]{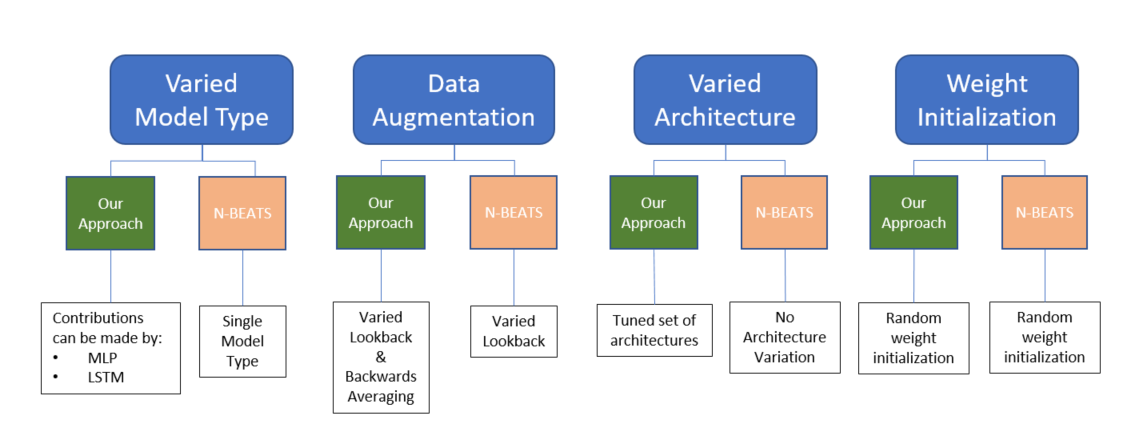}
    \caption{Example of how various ensemble diversity methods can be implemented. In this case between this work and the N-BEATS approach used by Stevenson et al. (2022).}
    \label{fig:Diversity Comparison Chart}
\end{figure}

With the basics of model ensembles discussed, it must be decided how best we approach forecasting $F_{10}$ and whether ensembles will be useful. To use neural networks for prediction of the proxy; we must decide what data to include, what base models to begin with, how these models perform, and whether using ensemble methods is advantageous. An overview of the methods used by our work are covered in Figure \ref{fig:Diversity Comparison Chart}.

Since an AR model only has previous values to use for forecasting, there may be difficulty in prediction using only previous values directly. An additional input known as back average is used to extract more information from the previous values,
\begin{equation}
    \overline{F}_{10_{B}} =\frac{1}{B} \sum_{j=1}^{B} (F_{10_{T-j}})
\end{equation}
where $B$ is the window of values to use for the backwards average.
By providing AR models with backwards averages a short-term trend may be identified and may guide the model towards a better average prediction. The sensitivity of model performance to this back average value will be investigated. Another method which considers the variation of lookback window, $L$, has been used as a method for promoting ensemble diversity \cite{Stevenson2022}, \cite{Oreshkin2019}, and \cite{AlShareef2010}. 
The sensitivity of model performance on lookback will also be investigated to determine the best lookback values to use during training.
\subsubsection{Manipulation of Learning Algorithm}
To promote further diversification in ensemble members, we consider manipulation of the learning algorithm by varying the model hyperparameters, an idea discussed by \cite{Sagi2018}. To accomplish this task, we consider manipulation via two methods.
First, we explore diversity through varied architecture using a hyperparameter tuner. To select skilled architectures, hyperparameter tuning was performed via KerasTuner, a hyperparameter optimization framework in the Keras Machine Learning API. A range of hyperparameters (example parameters seen in Table \ref{Tuning}) were considered, with KerasTuner outputting a list of models and hyperparameters with the best loss values. We consider these to be the best performing architectures for a given lookback and this is the method for generating ensemble members.

Next, we explore diversity via weight initialization. The initialization of weights in neural networks has a considerable impact on the local minimum that is found \citeA{Swirszcz2016}. In addition, \cite{AlShareef2010} determined that model performance can be improved by “just optimizing the initial random weights that the ANN starts with during training”. To improve model performance, we consider creating and training models several times per architecture. A similar method was used for creating diversity in ensemble members in \cite{Stevenson2022}. We consider the use of 10 models per architecture, initialized with a random weighting scheme.

\subsubsection{Hyperparameters and Best Model Selection}
During tuning, it is necessary to specify hyperparameter ranges to perform a search over. A Bayesian optimization scheme was used for selecting the best hyperparameters. We select the architectures that produce a minimal loss value on the validation set based on hyperparameter tuning. After tuning, the top 3 architectures are selected and saved to generate base ensemble members. These architectures are used to generate individual models whose weights are each randomly initialized. These randomly initialized models are then trained further with an early stopping criterion. After training, predictions on the test set occur and are saved to files individually. It should be noted that ensemble integration has not occurred yet and that the predictions are independent at this point.
\begin{table}[h]
 \caption{Tuning configurations to Generate ensemble members at each lookback for MLP and LSTM}
 \centering
 \small
  \label{Tuning}

 \begin{tabular}{l c c c}
  \multicolumn{4}{c}{\textbf{MLP}} \\
 \hline
  Tuner Option  & Choice & Parameter & Value/Range  \\
 \hline
   Scheme & Bayesian Optimization  &Number of Dense Layers &[1-8]  \\
   Total Trials & 100 &Dense Neurons & [4-256] \\
   Initial Points  &  5 & Activation & [relu,tanh,sigmoid]\\
   Repeats per Trial  &  3 & Optimizer& [adam, SGD]\\
   Minimization Parameter   &MSE   &Learning Rate& [.01, .001, .0001]\\
   Epochs  & 100  & Early Stopping Criteria  &    validation loss (MSE)  \\
   Early Stopping Patience & 30  &\\
 \hline
 \multicolumn{4}{c}{\textbf{LSTM}} \\
 \hline
  Tuner Option  & Choice & Parameter & Value/Range  \\
 \hline
   Scheme & Bayesian Optimization  &Number of LSTM Layers &[1-3]  \\
   Total Trials & 75 &LSTM Neurons & [4-128] \\
   Initial Points  &  5 &LSTM Activation & tanh\\
   Repeats per Trial  &  3 &Number of Dense Layers&[1-4]\\
   Minimization Parameter   &MSE   &Dense Neurons& [4-256]\\
   Epochs  & 50  &Dense Activations& [relu, tanh, sigmoid]\\
   Early Stopping Criteria  &    validation loss (MSE) &Learning Rate& [.01, .001, .0001]\\
   Early Stopping Patience & 10  &Optimizer& [adam, SGD]\\
 \hline
 \end{tabular}
 \end{table}

\subsubsection{Lookback and Combined Model Ensembles}

Tuner results from each architecture are used to create 10 random weight initialized copies and further training of models occur. The prediction of test set data is then made by each individual model and is saved for combination later. 
\begin{figure}[h]
    \centering
    \includegraphics[scale=0.75]{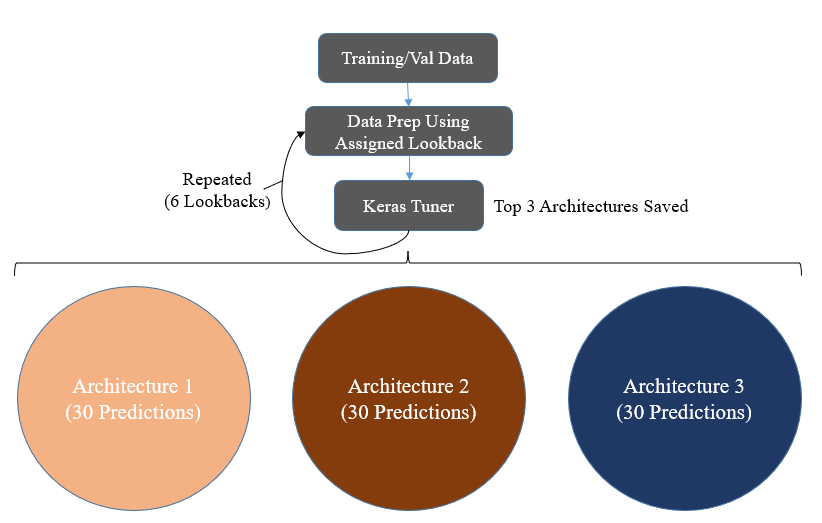}
    \caption{Predictions on test set are made by 30 models of each architecture for each lookback step and are saved for combination later. A total of 180 models are used for prediction.}
    \label{fig:MLE Architecture}
\end{figure}

Due to varied performance, it may be beneficial to consider using more than one model type at a time. For example, there may be periods of time where an MLP outperforms an LSTM (or other models). A hybrid ensemble approach like \cite{Phyo2021}, will be considered as well. In this work, we will evaluate the performance of combining the ensemble prediction across multiple model types, leveraging their differing skill. In the case of this work, we consider an unweighted average of predictions made by the MLP ensemble and LSTM ensembles.

These ensemble methods also provide a set of predictions and can be used to form a probabilistic distribution of predicted values. By evaluating the statistics across the distribution of ensemble member predictions, uncertainty in ensemble prediction can be quantified. Deterministic approaches lack the ability to provide an associated uncertainty with predictions. By providing both a predicted value and an associated uncertainty, proper response can be made by agencies that rely on a predicted $F_{10}$ value. It is also necessary to evaluate the robustness and reliability of the predicted uncertainty estimates, as can be seen in the previous work by \cite{Licata2022b}. This can be performed by calculating the confidence interval using the true data,
\begin{equation}
    CI = \Bar{x} \pm z*\frac{\sigma}{\sqrt{n}}
\end{equation}

where $\Bar{x}$ is the sample mean, $z$ is the confidence value given by , $\sigma$ is the sample standard deviation, and $n$ is the number of samples. To measure the "performance" of uncertainty, we calculate the cumulative percentage of true points capture in the ensemble uncertainty at a given point and compare it to the expected cumulative percentage, plotted as a calibration curve. A perfectly calibrated model would have curves that lie along a line of slope 1. If the curve lies above or below the line, then the ensemble is over or under predicting the uncertainty respectively. As discussed by \cite{Licata2022b}, we will also use the calibration error score (CES) metric. The CES gives a quantitative measure of how much error exists in the predicted uncertainty bounds by averaging the error of cumulative probability. Additionally, $\sigma$ scaling can be used to shift the uncertainty estimates based on performance on the validation data set. Essentially, this scaling uses the validation set to evaluate the predictions captured for a given CI and is used to scale the variance.
\begin{equation}
    \sigma_{S} = \sqrt{S}*\sigma = \sqrt{\frac{1}{N} \sum_{i=1}^{N} \sigma_{i}^{-2}*(y-\hat{y})^{2}} * \sigma ,
\end{equation}
where $S$ is the scaling factor , $N$ is the number of samples in the validation set, $\sigma_{i}$ is the sample standard deviation at step i, $\sigma_{S}$ is the scaled standard deviation and $(y-\hat{y})^{2}$ is the MSE of prediction.
To identify the biases and temporal uncertainty in predicted values, we evaluate the error statistics and compare using the binned results method presented in \cite{Licata2020}. By creating a set of bins of predicted values, the model uncertainty and biases at various solar activity levels can also be highlighted. A prediction is placed in a bin depending on the first predicted value and the binning values can be seen in Table \ref{Binning}.

\begin{table}[h]
 \caption{$F_{10}$ Binning used by Licata et al. (2020)}
 \centering
 \begin{tabular}{l c c}
\hline
 \label{Binning}
  Solar Activity Level  & Lower Bound [sfu] & Upper Bound [sfu] \\
 \hline
    Low & 0 & 75 \\
    Moderate & 75 & 150\\
    Elevated & 150 & 190 \\
    High & 190 & -- \\
\end{tabular}
 \end{table}

\section{Results}
\subsection{Input Sensitivity Study}
\begin{figure}[h]
  \centering
  \begin{minipage}[t]{0.5\textwidth}
    \includegraphics[width=\textwidth]{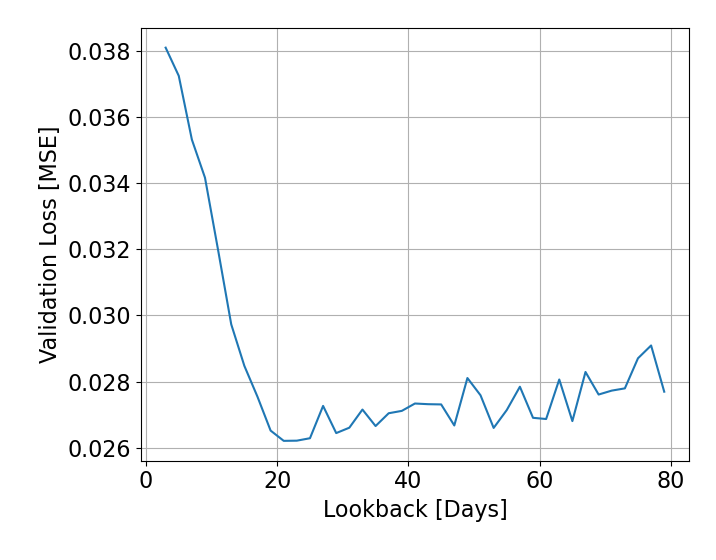}
    \caption{Averaged validation loss for a variety of architectures indicated a short term lookback between 2 and 3 weeks is better than very short or very long lookbacks.}
    \label{fig:Lookback Investigation}
  \end{minipage}
  \hfill
  \begin{minipage}[t]{0.50\textwidth}
    \includegraphics[width=\textwidth]{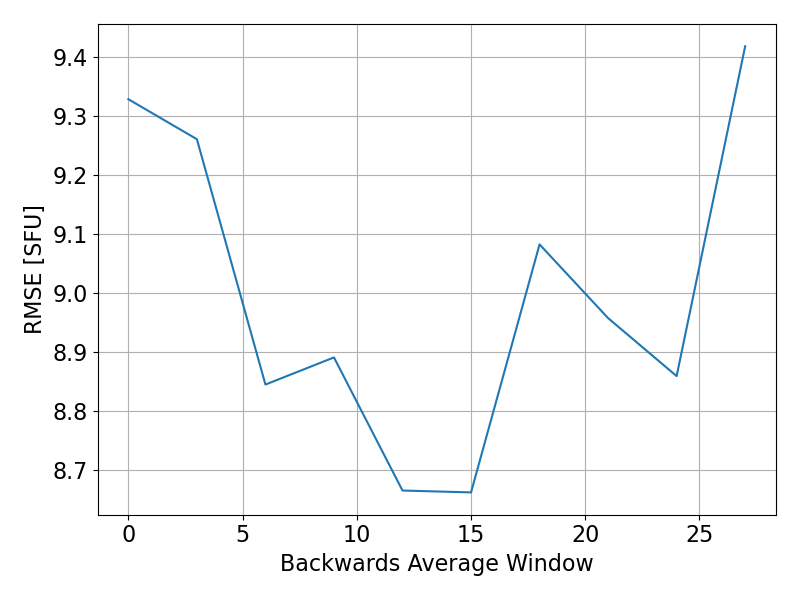}
    \caption{Training of models with different $B$ values indicates that short term backwards averaging helps but longer term averaging performed the same or worse than $B$=0.}
    \label{fig:BackAve}
  \end{minipage}
\end{figure}

In Figure \ref{fig:Lookback Investigation}, a range of lookback values were considered over 3 solar rotations (81 days). Random architectures were trained and evaluated on validation set for each lookback. This study resulted in a performance over lookback window and can be evaluated to identify important lookback values to include in the input for this work.
Based on the results from this analysis, we select a range of lookbacks $L = [7,10,13,16,19,22]$ days.

In Figure \ref{fig:BackAve}, we clearly see an increase in performance by including back average values during tuning/training. We select the best value from this sweep based on the metrics on the validation set to be used as an additional input to the neural network models. Based on the results of the backwards averaging study, we select a value for backwards average, $B$= 12 day.
\begin{figure}[!h]
    \centering
    \includegraphics[scale=0.50]{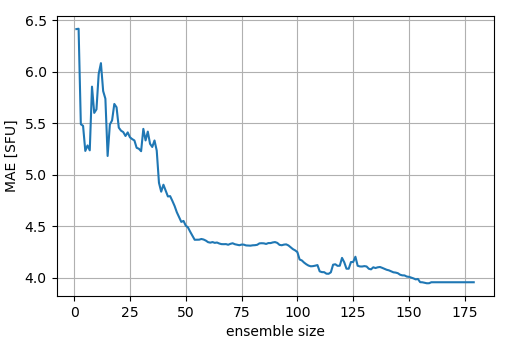}
    \caption{By incorporating more predictions, we are able to improve performance when compared to a single model.}
    \label{fig:Ensemble Size MLP}
\end{figure}

Using the optimal lookback and backwards average values, the performance of including additional models can be seen in Figure \ref{fig:Ensemble Size MLP}. This ensemble is created using MLP models with various lookbacks, random weighting, and architectures. In Figure \ref{fig:Ensemble Size MLP}, it can be seen that inclusion of ensemble members, lowers the error substantially. As seen in Figure \ref{fig:Horizon Metrics} the MAE decreases from 6.4 to 3.9 (approximately 48\%) from 1 member to 180 members, respectively. We also see a plateau and a lack of improvement at the end of the study, so 180 ensemble members are chosen.

\subsection{Predictions Compared Between Models}

\begin{figure}[h]
    \centering
    \includegraphics[width=\textwidth]{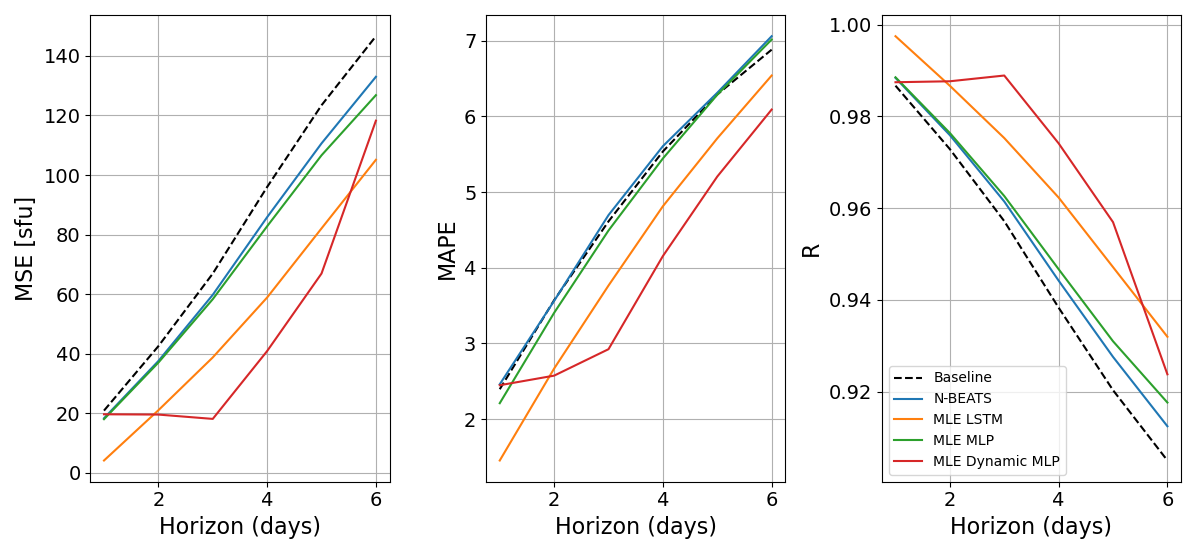}
    \caption{Evolution of metrics over forecast horizon indicate short term prediction is improved by ensemble methods.}
    \label{fig:Horizon Metrics}
\end{figure}

To compare model performances directly we must ensure that metrics are reported for the same test set data. We use the test data set presented in Figure \ref{fig:TrainTestSplit} to make predictions using a N-BEATS ensemble and a baseline model. 
When directly compared over the short term, the proposed ensemble methods outperform the baseline model in all cases, as can be seen in Figure \ref{fig:Horizon Metrics}, this indicates that ensemble methods are better at capturing non-linearity in the proxy. 
\begin{table}[!ht]
\centering
\caption{Relative metric comparison of Multi Lookback Ensemble (MLE) and Dynamic (D) methods with other forecasting methods. Metrics are scaled against the linear baseline, and averaged over forecast horizons. Lower error metrics and higher correlation metrics are preferred, with a value of 1 exhibiting the same performance as the linear baseline. The best performing values in each metric are highlighted in bold.}
\begin{tabular*}{\textwidth}{lcccc}
\label{Table}
 Model & & Relative Metric & &\\
                  \hline
                  \textbf{\small Training Set Sample (1964-1974)} & MSE   & RMSE& MAPE& R     \\
\hline
        N-BEATS & 0.815 & 0.809 & 0.972 & 1.001 \\ 
        MLP & 0.750 & 0.849 & 0.980 & 0.979 \\ 
        MLP$_{MLE}$ & 0.738 & 0.765 & 0.894 & 1.033 \\ 
        LSTM & 0.792 & 0.769 & \textbf{0.798} & 1.061 \\ 
        LSTM$_{MLE}$ & \textbf{0.535} & \textbf{0.762} & 0.894 & \textbf{1.064} \\ 
        MLP$_{MLE}$ + LSTM$_{MLE}$& 0.536 & 0.763 & 0.894 & 1.013 \\ 
        MLP$_{MLE,D}$ & 1.042 & 0.823 & 0.997 & 0.965 \\ 
        LSTM$_{MLE,D}$  & 0.985 & 0.876 & 0.980 & 0.994 \\ 
        MLP$_{MLE,D}$ + LSTM$_{MLE,D}$ & 0.844 & 0.764 & 0.892 & 1.061 \\ 
        MLP$_{MLE,D}$  + LSTM$_{MLE}$ & 0.555 & 0.769 & 0.856 & 1.090 \\ 
        MLP$_{MLE}$ + LSTM$_{MLE,D}$ & 0.544 & 0.771 & 0.853 & 1.012 \\ \hline
 \textbf{\small Validation Set (1994-2004)} & &  & &\\
\hline
        N-BEATS & 0.817 & 0.833 & 0.833 & 1.017 \\ 
        MLP & 0.766 & 0.807 & 0.823 & 1.019 \\ 
        MLP$_{MLE}$ & 0.756 & 0.802 & 0.808 & 1.019 \\ 
        LSTM & 0.770 & 0.809 & 0.821 & 1.021 \\ 
        LSTM$_{MLE}$ & 0.760 & 0.804 & 0.818 & 1.023 \\ 
        MLP$_{MLE}$ + LSTM$_{MLE}$  & 0.755 & 0.802 & 0.810 & 1.019 \\ 
        MLP$_{MLE,D}$  & 1.258 & 0.899 & 0.939 & 1.028 \\ 
        LSTM$_{MLE,D}$  & 1.287 & 0.908 & 0.955 & 1.028 \\ 
        MLP$_{MLE,D}$ + LSTM$_{MLE,D}$  & 1.266 & 0.899 & 0.928 & 1.028 \\ 
        MLP$_{MLE,D}$  + LSTM$_{MLE}$ & \textbf{0.462} & \textbf{0.662} & \textbf{0.695} & \textbf{1.037} \\ 
        MLP$_{MLE}$ + LSTM$_{MLE,D}$ & 0.585 & 0.696 & 0.701 & 1.031 \\ \hline
                   \textbf{\small Test Set (2006-2020)} & & \\
                   \hline
N-BEATS     & 0.874 & 0.913 & 0.901 & 1.009 \\
MLP             & 0.892 & 0.944 & 0.992 & 1.007 \\
MLP$_{MLE}$               & 0.866 & 0.930 & 0.975 & 1.008 \\
LSTM            & 0.607 & 0.772 & 0.883 & 1.019 \\
LSTM$_{MLE}$              & 0.544 &  0.725 & 0.816 & 1.021 \\
MLP$_{MLE}$  + LSTM$_{MLE}$         & 0.558 & 0.734 & 0.813 & 1.020 \\
MLP$_{MLE,D}$      & 0.574 & 0.742 &  0.807 & 1.025 \\
LSTM$_{MLE,D}$     & 1.703 & 1.167 & 1.317 & 1.005 \\
MLP$_{MLE,D}$ + LSTM$_{MLE,D}$  & 0.909 & 0.888 & 1.005 & 1.024 \\
MLP$_{MLE,D}$ + LSTM$_{MLE}$ & \textbf{0.442} & \textbf{0.659} &\textbf{0.759} & \textbf{1.029} \\
MLP$_{MLE}$ + LSTM$_{MLE,D}$  & 0.684 & 0.802& 0.901& 1.027 \\ 
\hline
\end{tabular*}
\end{table}

It can be concluded from Table \ref{Table} that the ensemble methods outperform persistence, the baseline, and N-BEATS for forecasting short term. It should be noted that dynamic prediction methods were found to improve performances in MLP models but hindered the predictions of the LSTM models. LSTMs, were found in the test case, to not benefit from the recursive prediction the same way MLPs did. This may be attributed to instabilities in predicted values, where large error predictions continued to be fed into the model, magnifying small errors. It should also be noted that the validation set had a higher solar maximum than the test set, and may have contributed to a few metrics being worse on the validation set.

It was also found that mixing prediction type helped significantly, using a mixed prediction (average prediction of MLP$_{MLE,D}$ and LSTM$_{MLE}$), we see an approximate 16\% MSE improvement over the single best non ensemble model on the test set. This may be due to the number of predictions averages, for the combined type ensembles, over 360 predictions are used, which may have resulted in a more accurate mean value. These results show that by using predictions made by an ensemble improve on the single model approaches. The relative Pearson correlation coefficient across various approaches reinforces that deep-learning approaches all out perform the baseline model. 

\subsection{Solar Activity Level and Error Statistics}
To identify uncertainty in prediction over forecast horizon we plot the Mean Error (ME) vs the horizon; the ME indicate the over/under predictions made by various models. This analysis uses the same dates for predictions as \cite{Licata2020} which is October 1$^{st}$ 2012 through December 31$^{st}$ 2018 and uses the binning criteria in Table \ref{Binning}. 
\begin{figure}[htbp]
    \centering
    \includegraphics[width=1.1\textwidth]{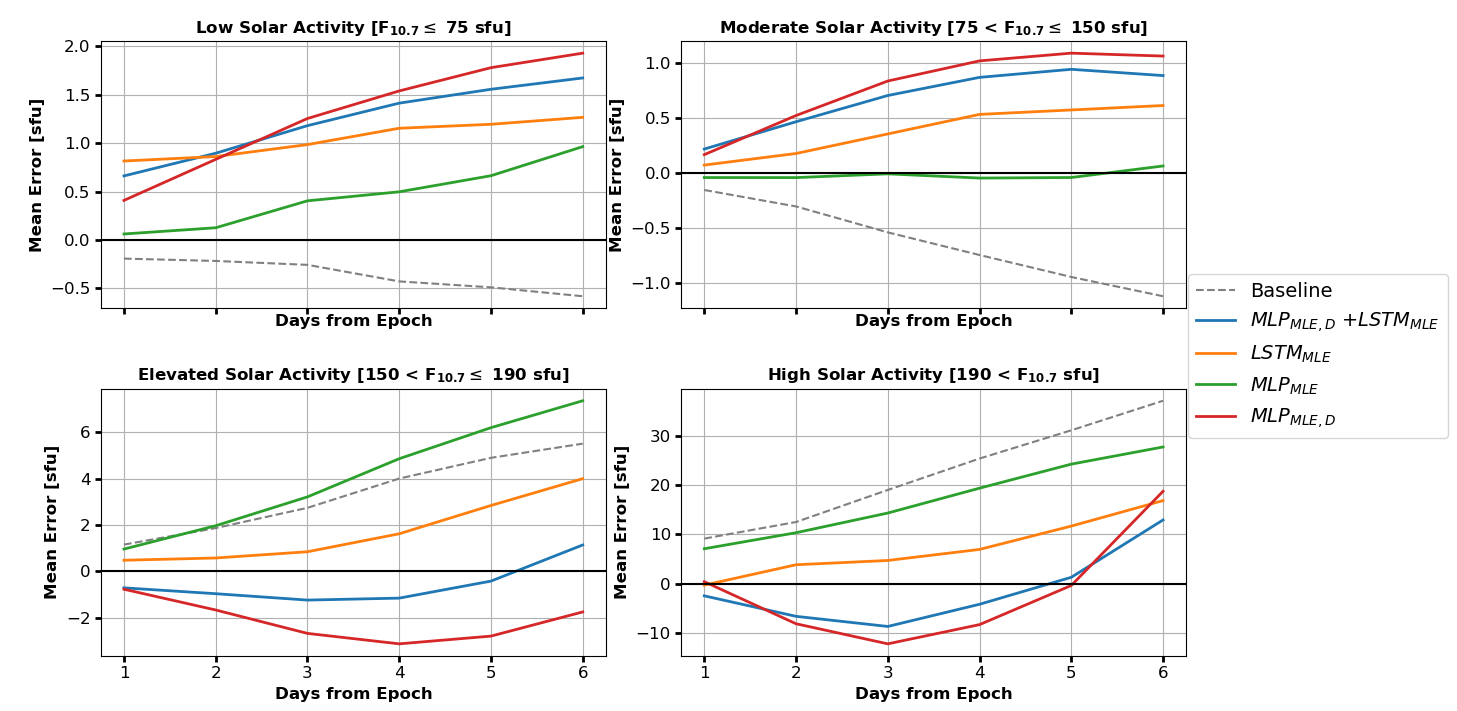}
    \caption{Bias of well performing models on test data indicates that ML methods provide much less biased predictions when solar activity is increased. The black line at zero indicates no bias.}
    \label{fig:Bias}
\end{figure}
In Figure \ref{fig:Bias}, it can be seen the predictions are less biased than the baseline method at high solar activity levels. At low activity, the biases are worse than the baseline, indicating a tendency to over predict by 1 SFU on average by the 6th day. At elevated and high activity levels, the dynamic prediction tended to under predict in some cases. We see that the ensemble that incorporates both multi step prediction and dynamic, made less biased predictions at both elevated and high activity levels but made more biased predictions at low activity levels. This indicates it may be useful in combining ensemble predictions in another way. The bias at low activity levels may be due to the training scheme; during training, the MSE loss function was minimized, penalizing larger errors greater than smaller errors. With large errors, training would skew the model to make better predictions at higher activity levels, to minimize the loss function. The error statistics at the high activity level are only analyzed for a small number of predictions (about 20 forecasts) and may not be statistically significant. The low, moderate, and elevated activity levels have approximately 500, 1400, and 200 forecasts respectively.

\begin{figure}[!h]
    \centering
    \includegraphics[scale=0.6]{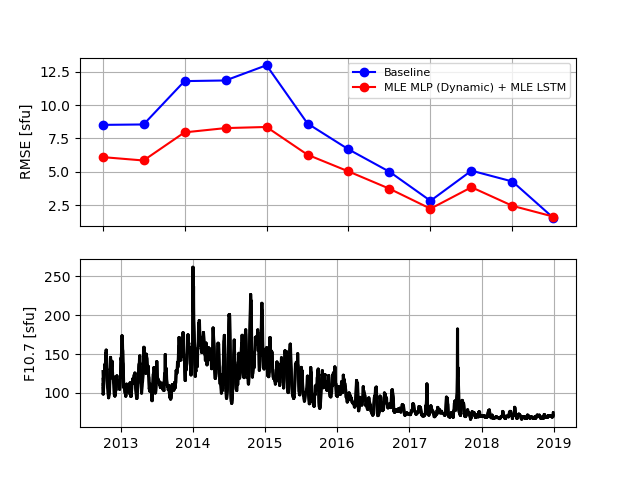}
    \caption{Top: 180 day averaged RMSE for forecast horizon of 6 days. Bottom: $F_{10.7}$ proxy changes with solar activity level over part of the test set.}
    \label{fig:SC Error Plot}
\end{figure}

Errors in prediction are suspected to be larger during periods of high solar activity. We break down the performance of the best ensemble approach and the baseline, along with the $F_{10}$ value from the maximum of solar cycle 24 to the minimum of solar cycle 25 . This is done in Figure \ref{fig:SC Error Plot} to highlight the high correlation between activity level and error. Our work is in agreement with the results seen in \cite{Licata2020} and \cite{Stevenson2022}; forecasting of $F_{10}$ is more challenging at higher solar activity levels. 

\subsection{Quantified Uncertainties}
The ensemble approach used in this work would allow a user to sample from a probabilistic range of values, rather than use a single deterministic value (like the baseline). By forecasting a range of $F_{10}$ values, a prediction contains an average and uncertainty bounds, which creates a more robust and operationally useful forecast. The averaged ensemble prediction should allow for a reduction in fluctuations predicted by individual models.

\begin{figure}[h]
    \centering
    \includegraphics[scale=0.4]{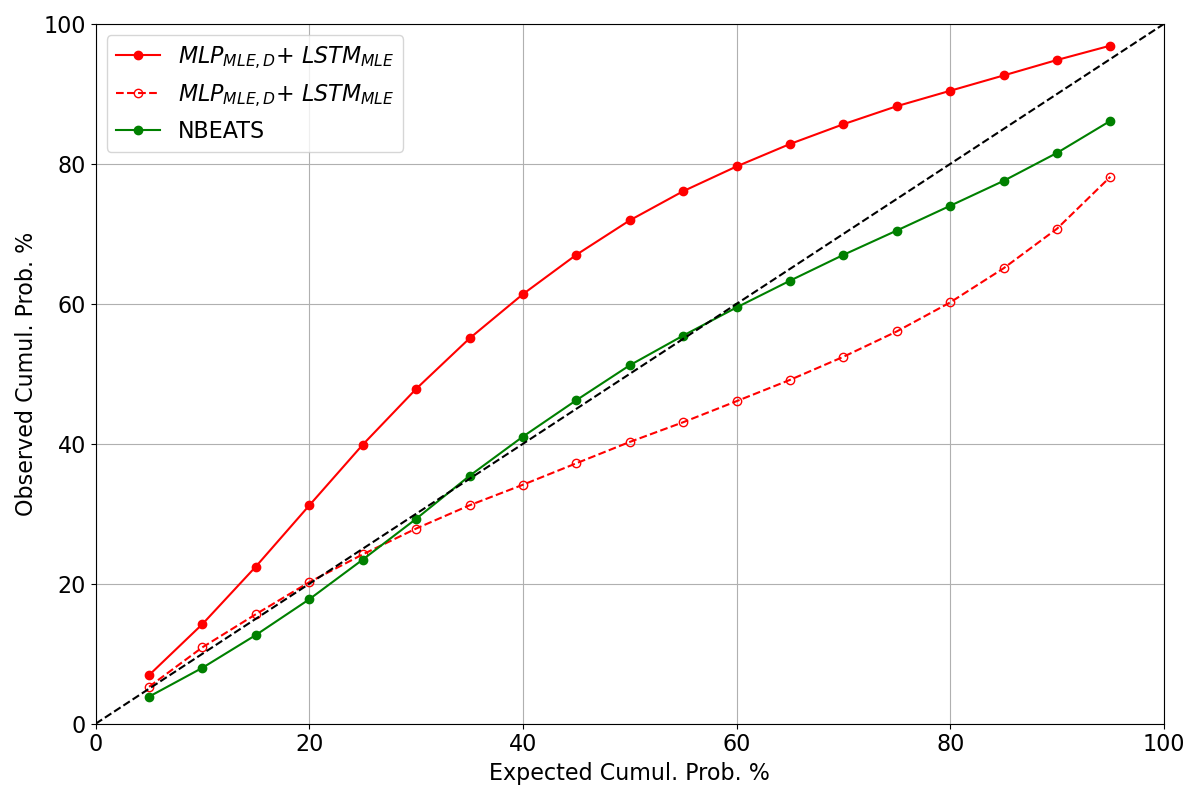}
    \caption{Test set calibration curves for ensemble predictions. Dotted line indicates $\sigma$  scaling has been applied.}
    \label{Calibration Curves}

\end{figure}
In Figure \ref{Calibration Curves}, we can see that models that have good metric scores are not necessarily the best in when predicted uncertainties are evaluated. The multi step prediction models tend to be under confident in the predicted uncertainty bounds and dynamic prediction causes the prediction to be over confident. The use of $\sigma$ scaling, or adjusting the uncertainty bounds based on validation performance is seen in the dotted lines in Figure \ref{Calibration Curves}. This scaling provides a better grouping of models along the 45$^{o}$ line, indicating a general improvement in uncertainty predictions after adjustments are made. It should be noted that \cite{Stevenson2022} varied the loss function used during ensemble member training and we did not. We believe that not enough evidence was given to support the argument that variation of the loss functions affected both uncertainty metrics and direct forecast metrics. A variation in loss function during training could potentially improve the CES metric, as seen in the NBEATS results in Table \ref{CE}. 

\begin{table}[h]
 \caption{Calibration Error Score (CES) of ensemble with good performance metrics as indicated by Table \ref{Table} and another ensemble approach by (Stevenson et al., 2022). Bold indicates best metric.}
 \centering
 \small
 \begin{tabular}{l c c c c c c }
\hline
 \label{CE}
  Model  & Test & Scaled Test  &Val.& Scaled Val.& Train  &  Scaled Train \\
    \hline
    MLP$_{MLE,D}$ + LSTM$_{MLE}$ &13.4 & 9.8 & 14.34 & \textbf{5.22} & 18.20 & 14.51\\
    NBEATS & \textbf{2.87} & \textbf{2.38} & \textbf{12.52} & 8.65 & \textbf{14.79} & \textbf{11.8}\\
 \hline
\end{tabular}
 \end{table}
It is interesting that a model with worse performance metrics (error) makes better uncertainty bound predictions, as seen in Table \ref{CE}. One possibility for the NBEATS performance may be due to the diversity through varied loss function. By varying loss function NBEATS may be more suited for uncertainty estimation than direct value prediction and indicates that without scaling, the NBEATS approach is well calibrated. Although, after scaling was applied based on validation performance, we improve the uncertainty estimation significantly for our ensemble methods, reducing the CES by 4\% and 9\% on the test and validation sets respectively, while using the $MLP_{MLE,D}$ + $LSTM_{MLE}$ ensemble. This indicates that $\sigma$ scaling can be useful for making more robust and reliable uncertainty predictions when validation data is available. It should also be noticed that uncertainty estimation on the test set is better than both validation and training sets, which may be attributed to the difference in solar activity levels (higher maximums), see Figure \ref{fig:TrainTestSplit}. 
\section{Summary and Conclusions}

In this work, the ability of neural network ensemble methods for predicting the $F_{10}$ proxy were investigated and results were compared against an operational linear model. In this work, input data manipulation was considered. A variation of lookback window indicated that short term lookbacks between 2 and 3 weeks were preferred and the inclusion of a backwards averaged value of about 14 days showed a 5\% improvement over no backwards averaging. Several lookback windows were considered and our results agree with the work done by \cite{Stevenson2022}, indicating variation of lookback provides greater model performance. Neural network ensembles allowed us to improve on the baseline performance metrics in nearly all ensemble approaches used and also provides a measure of uncertainty in predicted values due to the distribution of predicted values. An investigation into ensemble member size also indicated that a 48\% improvement of MAE was seen between a single model vs an ensemble of size 180.

Model ensembles were evaluated over a six day forecast horizon and compared directly to the baseline approach using both multi step and dynamic forecasting methods. Dynamic forecasting methods did not show large improvement on their own. Applying a mixture of multi step and dynamic methods showed the best results, specifically using a combination of dynamic MLP and multi step LSTM. This work indicated a substantial improvement on the short term forecasting of the $F_{10}$ solar radio flux proxy when compared against SET's linear prediction algorithm; specifically when using neural network ensembles. The best ensemble methods improved the average relative MSE between a 45-55\% and showed an improvement in correlation. Additionally, model performance was evaluated at various solar activity levels and indicated that ensemble approaches were more biased at low conditions, but equivalent or better at moderate, elevated, or high. These results may indicate that model loss values are punished at larger solar activity levels and these effect should be further investigated.

It is important to understand the spread of predicted values. The baseline and other linear methods lack the ability to provide a distribution of predictions, yet  ensemble predictions can be used to generate multiple predictions. Using a distribution of predicted values, calibration curves and CES metrics can be investigated. By analyzing uncertainty on a validation set, ensemble methods are able to provide the tendency of a model to over or under predict, and steps like $\sigma$ scaling can be taken to address those tendencies and increase the robustness of uncertainty estimates. In this work, it was shown that scaling allowed for a 9\% improvement in uncertainty estimation on validation data. By harnessing the ability of ensembles to make a probabilistic forecast, more robust and reliable satellite state estimation can be performed. 

Neural network ensembles are a powerful tool that should further be used in the field of space weather forecasting. As more research is done in the field, prediction methods for solar and geomagnetic indices and proxies can most likely be improved. In addition to direct averaged forecasts, we generated a distribution of predicted values which can lead to future investigations on space weather driver and model uncertainty effects. This work will hopefully help lead to future improvements made in prediction of not only $F_{10}$ but many of the other solar and geomagnetic indices. Data splitting into training and validation data was performed to be consistent with the previous work done by \cite{Stevenson2022}, but this validation set may not include values large and small enough to correctly validate the training data. It should be noted that by manipulating the validation set (by a method such as K-Fold cross validation) may further improve results.

\section{Limitations and Future Work}
To improve forecasting of $F_{10}$ using only historical values and NN ensembles, investigation into more advanced ensemble training or integration approaches are necessary. Greater ensemble diversity, specifically through varied loss function, should be investigated further. Ensemble member generation may also be a key area for future work, as methods such as boosting or stacking could develop members that are more skilled in various areas. In addition, the work by \cite{Liu2000} discussed training of ensemble members simultaneously (evolutionary ensembles) may allow for more diverse members that are still individually skilled by enforcing negative correlation during training.

Although this work shows improvement on forecasting of $F_{10}$ for short periods, longer term proxy forecasting is necessary. The models and approaches presented in this work may be suited for short term forecasts, but may not perform as well as other approaches for longer term predictions. One possible approach that could be taken to further improvements may involve the addition of a solar cycle variable. A model may learn better if it is "told" roughly what solar activity level is expected. It may be necessary to develop models and approaches that are used for different lengths of forecast. There is only so much information that can be gathered from previous values alone, even using neural networks. The addition of solar disk images, suggested by Benson et al. may be critical in improving longer term forecasts, due to the time lag between solar activity and thermosphere response. To improve forecasting of $F_{10}$ in the long term, it is necessary to improve our understanding of the Sun's physical and dynamical processes. With a better understanding of the Sun's physics, prediction of $F_{10}$ should be improved further.

In future works, it may be critical to evaluate also the "time-lag" of predicted values when using AR models. If a solar storm were to occur and a prediction of the proxy is made only after the event, then the prediction is meaningless and would not help in the real world. We recommend that future forecasting evaluations consider in addition to metrics, an analysis of delayed response of model. Similarity between signals at different times can be accomplished with an algorithm such as Dynamic Time Warping (DTW) and may be useful in quantifying temporal delay. It is also noted, when the only input is previous lookback values, this response is typical and may not be avoided completely. This may be seen in operations when forecasts are sampled from the ensemble predictions.

\section{Open Research}
SET proprietary data for this research are not made publicly available since they reside on operational servers run for the sole benefit of the USAF. Data are provided courtesy of Space Environment Technologies, 2019. These data have been provided to West Virginia University with license to use for scientific research.

\section{Data Availability}

The results presented in this document rely on data collected by the Solar Radio Monitoring Program (https://www.spaceweather.gc.ca/forecast-prevision/solar-solaire/solarflux/sx-en.php) operated by the National Research Council and Natural Resources Canada. These data are available at https://www.spaceweather.gc.ca/forecast-prevision/solar-solaire/solarflux/sx-5-en.php. These data were accessed via the LASP Interactive Solar Irradiance Datacenter (LISIRD) (https://lasp.colorado.edu/lisird/).
\\

\acknowledgments

JDD and PMM greatly acknowledge support under NSF ANSWERS award \# 2149747 (sub award to West Virginia University from Rutgers University).

The authors declare that they have no known competing
financial interests or personal relationships that could have
appeared to influence the work reported.


%
%



\bibliography{probabilistic_solar_proxy_forecasting_with_neural_network_ensembles}
%
%
%
%
%
\end{document}